\documentclass[aip,jmp,amsmath,amssymb,
reprint%
]{revtex4-1}

\usepackage{graphicx}
\usepackage{dcolumn}
\usepackage{bm}
\usepackage[english]{babel}
\usepackage[utf8]{inputenc}
\usepackage{anysize}
\usepackage{graphicx}
\usepackage{color}
\usepackage{hyperref}
\usepackage{bm}
\usepackage{amsmath,amssymb}

\begin{document}


\title{Optothermal nonlinearity of silica aerogel}

\author{Maria Chiara Braidotti}
    \affiliation{Institute for Complex Systems, National Research Council (ISC-CNR), Via dei Taurini 19, 00185 Rome (IT).}
		\affiliation{Department of Physical and Chemical Sciences, University of L'Aquila, Via Vetoio 10, I-67010 L’Aquila (IT).}
    \email{mariachiara.braidotti@isc.cnr.it}
\author{Silvia Gentilini}
    \affiliation{Institute for Complex Systems, National Research Council (ISC-CNR), Via dei Taurini 19, 00185 Rome (IT).}
		
\author{Adam Fleming}
    \affiliation{School of Physics and Astronomy (SUPA), University of St Andrews, North Haugh, St Andrews KY16 9SS (UK).}
\author{Michiel C. Samuels}
    \affiliation{School of Physics and Astronomy (SUPA), University of St Andrews, North Haugh, St Andrews KY16 9SS (UK).}
\author{Andrea Di Falco}
    \affiliation{School of Physics and Astronomy (SUPA), University of St Andrews, North Haugh, St Andrews KY16 9SS (UK).}		
\author{Claudio Conti}
    \affiliation{Institute for Complex Systems, National Research Council (ISC-CNR), Via dei Taurini 19, 00185 Rome (IT).}
    \affiliation{Department of Physics, University Sapienza, Piazzale Aldo Moro 5, 00185 Rome (IT).}
    \homepage{http://www.complexlight.org}

\date{\today}

\begin{abstract}
We report on the characterization of silica aerogel thermal optical nonlinearity, obtained by z-scan technique. The results show that typical silica aerogels have nonlinear optical coefficient similar to that of glass $(\simeq 10^{-12} $m$^2/$W), with negligible optical nonlinear absorption. The non\-li\-near coefficient can be increased to values in the range of $10^{-10} $m$^2/$W by embedding an absorbing dye in the aerogel. This value is one order of magnitude higher than that observed in the pure dye and in typical highly nonlinear materials like liquid crystals.  
\end{abstract}

\maketitle

In the recent years, there has been growing interest in studying soft-colloidal matter and complex materials for their properties and possible applications in biophysics and photonic technologies.\cite{Likos01,Khoo95,Conti06,Lee09,DelRe11,Ghofraniha09,Ashkin82} In particular, their nonlinear optical properties are promising for fundamental physical studies, imaging and sensing. Despite these encouraging features, they present scattering losses and limits in the high energy regime due to the fact that thermal phenomena as diffusion and convection, which usually cause the destruction of the sample, are present.\cite{ContiDelRe10} \\
In this framework, nanoporous Silica Aerogel (SA) has attracted great interest\cite{AreogelBook} for its physical, chemical and mechanical characteristics. This solid-state material exhibits strong nonlinearities and low thermal conductivity with limited optical scattering losses. Its porous structure hampers convection by trapping the gas molecules within a large number of nano-cavities. This feature is called Knudsen effect\cite{KnudsenBook} and makes Aerogel an excellent thermal conductive insulator, also allowing steep temperature gradient profiles. Because of these characteristics, SA is a good candidate for very high optical power applications.  \\
Recently, physicists reported on SA nonlinear optical response.\cite{Seo2003,Seo2006} Thanks to a z-scan measurements, performed by using impulsed lasers at $800$nm and $532$nm wavelength, they found a nonlinear refractive index of electronic nature respectively of $-1.5 \times 10^{-15}$ m$^2/$W and $-4.0 \times 10^{-14}$ m$^2/$W. These values have been questioned by other measurements which state that they are too high. \cite{Birks10,Birks10_02}\\
Further studies have demonstrated shock waves excitation in SA due to the presence of a optothermal induced nonlinearity and resistance to highly focused laser beams.\cite{Gentilini:14} In these regards, quantifying the SA thermal nonlinearity and its temporal response is extremely relevant.\\
In the following, we report on an experimental investigation of thermal nonlinear index of refraction $n_2$ for SA samples. We first investigate the nonlinear response in time as a function of power, that demonstrates the thermal origin. Than, through z-scan technique, we measured the $n_2$ coefficient.\\

\begin{figure}[h!]
	\begin{center}
		\includegraphics[scale=0.22]{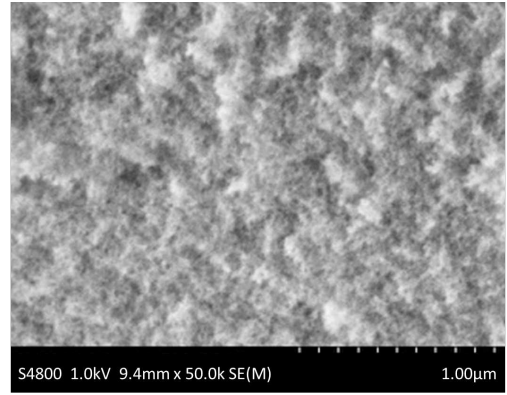}
	\end{center}
		\caption{SEM image of the SA sample.}
	\label{sa_photo}
\end{figure}

Silica Aerogel samples were synthesized by the coauthors by a base catalyzed polymerization process using tetramethylorthosilicate (TMOS) as the precursor \cite{Rao1998}. 
For a typical sample with a total volume of $1$ml, tetramethyl orthosilicate (TMOS; $400\mu$l, $2.71$mmol, $1$eq.) 
was diluted with methanol (MeOH; $400\mu$l, $9.89$mmol, $3.65$eq.).
After shaking the jar for $30$ seconds, an ammonium hydroxide solution (NH4OH (aq); $200\mu$l, $78.3$mM solution, $0.016$mmol, $0.006$eq.) 
was added and the mixture was shaken for $60$ seconds before pouring it into a PMMA cuvette mould, with a square cross section of $1$cm$^2$. 
The mould was sealed to prevent drying of the sample by evaporation of the methanol during the gelation. 
The formed alcogel was then allowed to set for approximately $1.5$ to $2$ hours before submerging the PMMA cuvette in acetone for solvent extraction. 
The PMMA cuvette dissolves in the acetone and by refreshing the acetone se\-ve\-ral times over a $4$ days period, the alcogel was puri\-fied and chemical impurities were removed. 
The alcogel was then dried by transferring it into a cu\-stom made supercritical point dryer to replace the acetone inside the gel with liquid $CO_2$, 
which was subsequently evaporated slowly in its supercritical phase, obtaining pristine SA samples.
Figure \ref{sa_photo} shows a SEM image of the SA sample used.
The matter density and the linear refractive index of the samples used are respectively $\rho = 0.215$ g$/$cm$^3$ and 
$n_0 = 1.074$: these parameters participate in determining the strength of scattering.

\begin{figure}[h!]
	\begin{center}
		\includegraphics[scale=0.35]{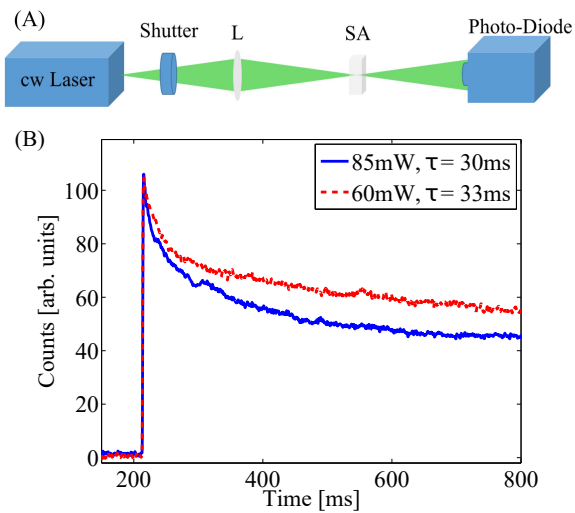}
	\end{center}
		\caption{(A) Sketch of the experimental setup used to perform time-dependent response of the SA sample. (B) Oscilloscope signal for two different beam powers $P=60$mW and $85$mW. The legend gives the measured decay times of different powers. Data have been scaled in order to allow a better view of the differences in the decay rates.}
	\label{panel3}
\end{figure}

In order to verify the thermal origin of the non\-li\-nea\-ri\-ty of the sample, we measure the non\-li\-near time-dependent response. 
We used a continuous-wave (CW) laser beam at wavelength $\lambda=532$nm, controlled in time by a beam shutter. 
The laser beam is focused in the SA sample with a focus size $\sim 10 \mu$m and a central section of the transmitted signal is acquired through a photo-diode connected to an oscilloscope (Fig.\ref{panel3}A).
Fi\-gure \ref{panel3}B reports the photon counts collected by the oscil\-lo\-scope for two different laser powers $P=60$mW and $85$mW. We observe that the beam intensity has different time decays at the two different powers consistently with their non\-linear nature. 
The decay rates, obtained fitting these curves, are $\tau_{60mW}=(33\pm1)$ms and $\tau_{85mW}=(30\pm1)$ms as reported in the inset of Fig.\ref{panel3}B. The decay rates are of the order of few milliseconds, as expected by a nonlinearity of thermal origin. \cite{BoydBook} Indeed, the decrease in transmitted intensity with time is due to the presence of a defocusing thermal nonlinearity, whose measurements are reported afterwords. When the thermal nonlinearity is excited, light is subject to a defocusing, i.e., the intensity decreases locally. This is coherent with measurements in Fig. \ref{panel3}B, which show intensity exponential decays exp$(-t/\tau)$ in the region of light covered by the photo-diode.\\

\begin{figure}[t!]
    \centering
			\includegraphics[scale=0.28]{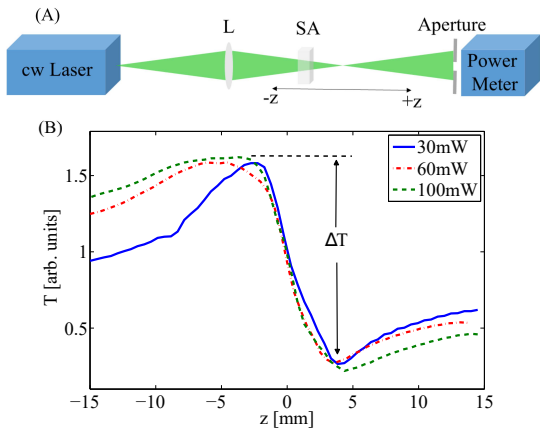}
		\caption{(A) Sketch of the experimental setup used to perform z-scan measurement. (B) Normalized transmittance of SA for different laser power $P=30$mW, $60$mW and $100$mW.}
		\label{panel1}
\end{figure}

In order to quantify the thermal nonlinear index of refraction $n_2$ of the sample, we used the z-scan technique\cite{Bahae90,Sheik-bahae:89,Bahae1991}. This consists in measuring the transmitted intensity through a finite aperture placed in the far field of a laser beam propagating through the sample, which is moved across a lens focal plane (see Fig.\ref{panel1}A). If the sample is sufficiently thin, i.e. it is smaller than the diffraction length, it behaves like a lens with variable focal length. The transmitted intensity as a function of the z-position of the sample presents a peak and a valley (see Fig.\ref{panel1}B).
The transmission, $T$, is calculated by normalizing the output power $P_{out}$ to its mean value.
The peak-valley difference $\Delta T$ is proportional to the phase shift induced by the presence of the sample, and hence to its nonlinear index of refraction, as shown in the following expression:\cite{Bahae90}

\begin{equation}
n_2=\frac{\Delta T}{0.406 (1-S)^{1/4} k L_{eff}I_0}
\label{n2}
\end{equation}
where $k$ is the wave vector and $I_0$ is the intensity at the focal point, $S$ is the linear transmittance of the far field aperture and $L_{eff}$ is the effective thickness of the sample.
The occurrence of the peak before the valley and viceversa depends on the sign of the nonlinearity.\\
 
\begin{figure}[t!]
    \centering
			\includegraphics[scale=0.38]{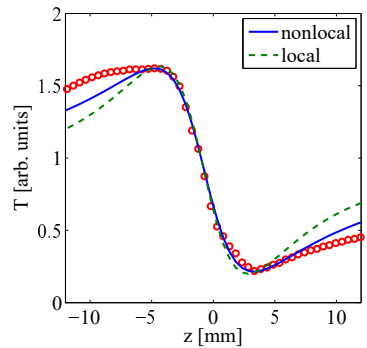}
		\caption{Local (dashed-green) and nonlocal (continue-blue) fit of the normalized transmittance of SA for laser power $P=100$mW.}
		\label{panel5}
\end{figure}

The  experimental setup is illustrated in Fig.\ref{panel1}A. As in the previous measurement, we used a CW laser beam at $532$nm wavelength. The sample is placed on a translation stage, which changes its position around the lens focus. We record the power transmitted by the aperture by a power meter. 
Figure \ref{panel1}B shows the normalized transmission curves obtained for the SA sample at different laser powers $P=30$mW, $60$mW and $100$mW. The deviation of curves in Fig. \ref{panel1}B from the expected lorentzian behavior is due to the presence of nonlocality.\cite{Ortega:14,Lara:07,Ramirez:10,DavilaPintle:13} Indeed, in a purely refractive medium nonlocality tends to deepen the valley and suppress the peak. \cite{Vaziri2013}
Furthermore, nonlocality acts on the transmission function broadening its tails (see Fig.\ref{panel1}B). Figure \ref{panel5} shows both local and nonlocal fits of the normalized transmission curves at $P=100$mW. The nonlocal fitting function used is 
\begin{equation}
T(z,\Delta\phi)=1-\frac{4m\Delta\phi x}{[x^2+(2m+1)^2](x^2+1)} 
\label{fit}
\end{equation}
where $m$ is the nonlocal coefficient, $\Delta\phi$ the nonlinear phase change induced by the material during the z-scan, and $x$ is the ratio between the propagation distance $z$ and the Rayleigh length $z_0$. The local case can be obtained putting $m=1$ in Eq. (\ref{fit}). \cite{Vaziri2013,Vaziri2013_02}
As shown in Table {\ref{table3}}, the value of the nonlinear refractive index $n_2$ does not change significantly due to the fitting model used.

\begin{table}[h!]
\centering
     \begin{tabular}{  c | c | c }
     $n_2$ [m$^2/$W] $\times10^{-13}$  & $m$  & fitting model\\ \hline
     $-4.0 \pm 0.2 $ & $1$ &  Local\\ 
     $-6.0 \pm 0.3 $ & $3.2 \pm 0.2$ &  Nonlocal\\ 
     \end{tabular}
	\caption{$n_2$ values for different fitting model at $P=100$mW. Uncertainties are obtained from the fitting procedure.}
	\label{table3}
\end{table}

We measured the nonlinear absorption opening all the aperture in Fig. \ref{panel1}A, not finding any measu\-rable value of the nonlinear absorption coefficient $\beta$. This feature shows that SA is an optimal sample for thermal nonlinear studies since it does not exhibit significant nonlinear losses. \\   
Repeating this measurement in different points of the sample we found that there is not a substantial dependence of the nonlinear refractive index $n_2$ value from the position in the sample. Hence we conclude that the sample is homogeneous, and the $n_2$ measurements are not affected by particular disorder configuration or scattering.
The retrieved values of the defocusing nonlinear refractive index are reported in Table \ref{table1}. Uncertainties are calculated from the different measurements at different points of the sample. As shown in Table \ref{table1}, the nonlinear refractive index is slightly dependent by the beam input power. This is due to the fact that the SA nonlinearity is not high, and hence the $n_2$ variation as a function of power is not evident. Despite this, we state that the sample has a defocusing nonlinearity.

\begin{table}[h!]
\centering
     \begin{tabular}{  c | c }
     P [mW]   & $n_2$ [m$^2/$W] $\times10^{-12}$ \\ \hline
     30       & $-1.23\pm0.13$   \\ 
     60       & $-0.67\pm0.05$   \\ 
     100      & $-0.41\pm0.01$   \\ 
     \end{tabular}
	\caption{$n_2$ values for different laser powers $P$ measured in different points of the sample. The sample disorder is homogeneous and $n_2$ measurements are not affected by disorder or scattering.}
	\label{table1}
\end{table}

\begin{figure}[h!]
	\begin{center}
		\includegraphics[scale=0.4]{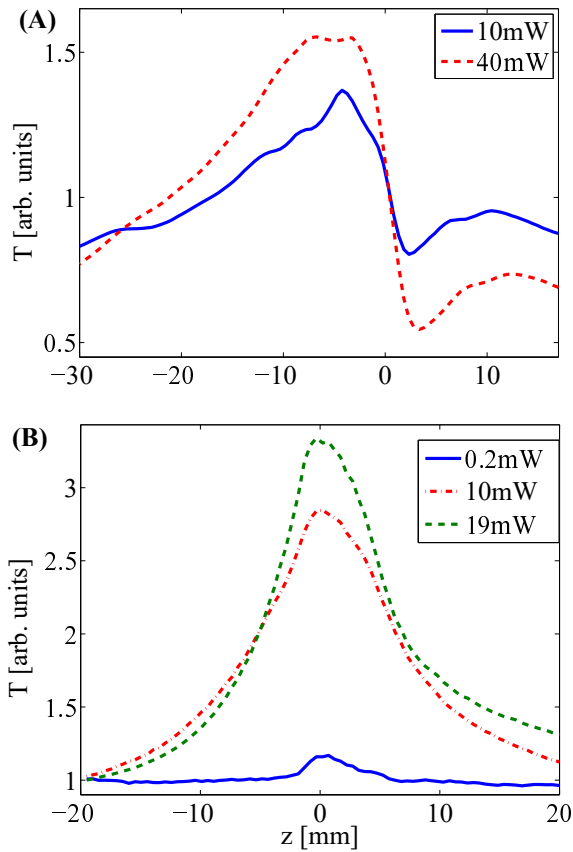}
	\end{center}
		\caption{(A) Normalized transmittance of the RhB-sample at two different powers $P=10$mW and $40$mW. From this we obtained the $n_2$ values respectively of: $n_2=(-3.4\pm0.2)\times10^{-10}$m$^2/$W and $(-1.4\pm0.1)\times10^{-10}$m$^2/$W. (B) Measurement of $\beta$ of the RhB-Aerogel Sample for different powers: $P=0.2$mW, $10$mW and $19$mW. We find $\beta=(4.0\pm0.2)\times10^{-4}$m$/$W, $(6.8\pm0.3)\times10^{-5}$ m$/$W and $(4.5\pm0.2)\times10^{-5}$m$/$W.}
	\label{panel2}
\end{figure}

In order to increase the nonlinearity we added Rhodamine B (RhB) to the SA sample, before the gelation stage, using a solution tetramethyl rhodamine iso-thiocyanate in methanol (TRITC; $2.7~mM$ in MeOH; $100 \mu l$, $0.27 \mu mol$). RhB is a dye and has a high nonlinear refraction index ($n_2\simeq-10^{-11}$m$^2$/W for an aqueous solution of RhB at a concentration of $0.6$mM)\cite{Ghofraniha07}. Figure \ref{panel2}A shows the normalized transmittance of this sample for different input beam powers $P=10$mW and $40$mW. We found an increase in the $n_2$ value with respect to the not-dyed SA sample: $n_2=(-3.4\pm0.2)\times10^{-10}$m$^2/$W and $(-1.4\pm0.1)\times10^{-10}$m$^2/$W respectively. The curve asymmetry is due to the pre\-sence of nonlinear absorption\cite{DavilaPintle:13} (Fig.\ref{panel2}A), while the nonlocality enhances the peak with respect to the valley\cite{Ortega:14}. Figure \ref{panel2}B reports the measurement of the nonlinear absorption $\beta$ which is found to be $(4.0\pm0.2)\times10^{-4}$m$/$W, $(6.8\pm0.3)\times10^{-5}$m$/$W and $(4.5\pm0.2)\times10^{-5}$m$/$W at $P=0.2$mW, $10$mW and $19$ mW. The thermodiffusion of RhB provides a measurable value of $\beta$: in the high intensity region, SA scaffold suffers a thermal expansion increasing the fraction of air volume in the lightened region. This phenomenon causes a reduction in the refractive index (defocusing effect) and in the amount of absorption (saturable absorption effect), and hence an increase in the transmission function. The broadening of the transmission function in Fig. \ref{panel2}B os related to nonlocality. 
We remark that $n_2$ and $\beta$ dependence on the power shows evidence of higher order nonlinear effects, which we ascribe to reversible structural deformations of the SA scaffold outlined before. 

\begin{table}[h!]
\centering
     \begin{tabular}{  c | c }
     Material   & $|n_2|$ [m$^2/$W]  \\ \hline
     Liquid Crystals\cite{GOMEZ2003}       & $10^{-11}$   \\ 
     RhB at $0.6$mM\cite{Ghofraniha07}       & $10^{-11}$   \\ 
     Glass\cite{dancus2007z} & $10^{-12}$ \\
		\end{tabular}
	\caption{Thermal $n_2$ values for various materials.}
	\label{table2}
\end{table}

The values of $n_2$ obtained for SA are in agreement with thermal nonlinearities reported in literature as can be seen in Table \ref{table2}. Adding RhB strongly enhances the nonlinear response.\\

In conclusion, we measured the nonlinear response of Silica Aerogel confirming its thermal origin and the fact that disorder does not hamper the nonlinear action: the strong nonlinearity of SA is comparable with the highest reported and can be enhanced by adding Rhodamine or other light absorbing dyes. 
We also find evidence of higher order nonlinear effects witnessed by the dependence of the nonlinear refraction and absorption on the input power. Thanks to its impressive properties Silica Aerogel can be exploited in several application as high power lasers or optical limiters.\\

We acknowledge support from EPSRC (EP/J004200/1 and EP/M508214/1), the Templeton Foundation (grant number 58277) and the ERC project VANGUARD (grant number 664782). The research data supporting this publication can be accessed at http://dx.doi.org/10.17630/0c377d2f-efa5-4228-9b73-5b1f36a8365e.


\end{document}